\documentclass[msmath,amssymb,prl,twocolumn,epsfig,showpacs,bibliography,lengthcheck,superscriptaddress]{revtex4-2}

\usepackage{graphicx}% Include figure files
\usepackage{dcolumn}% Align table columns on decimal point
\usepackage{bm}% bold math
\usepackage{hyperref}% add hypertext capabilities
\usepackage{epstopdf}
\usepackage{subcaption}
\usepackage{caption}
\usepackage{amsmath}
\usepackage{mathrsfs}
\usepackage{color}

\begin{document}

\title[\empty]{Search for spin-dependent gravitational interactions at the Earth range}

\author{S.-B. Zhang}
\author{Z.-L. Ba}
\author{D.-H. Ning}
\affiliation{CAS Center for Excellence in Quantum Information and Quantum Physics, School of Physical Sciences, University of Science and Technology of China, Hefei 230026, China}
\author{N.-F. Zhai}
\affiliation{Department of Precision Machinery and Precision Instrumentation, Key Laboratory of Precision Scientific Instrumentation of Anhui Higher Education Institutes, University of Science and Technology of China, Hefei 230027, China}
\author{Z.-T. Lu}
\email{ztlu@ustc.edu.cn}
\affiliation{CAS Center for Excellence in Quantum Information and Quantum Physics, School of Physical Sciences, University of Science and Technology of China, Hefei 230026, China}
\affiliation{Hefei National Laboratory, University of Science and Technology of China, Hefei 230088, China}
\author{D. Sheng}
\email{dsheng@ustc.edu.cn}
\affiliation{Department of Precision Machinery and Precision Instrumentation, Key Laboratory of Precision Scientific Instrumentation of Anhui Higher Education Institutes, University of Science and Technology of China, Hefei 230027, China}
\affiliation{Hefei National Laboratory, University of Science and Technology of China, Hefei 230088, China}
\date{\today}

\begin{abstract}
Among the four fundamental forces, only gravity does not couple to particle spins according to the general theory of relativity. We test this principle by searching for an anomalous scalar coupling between the neutron spin and the Earth gravity on the ground. We develop an atomic gas comagnetometer to measure the ratio of nuclear spin-precession frequencies between $^{129}$Xe and $^{131}$Xe, and search for a change of this ratio to the precision of 10$^{-9}$ as the sensor is flipped in the Earth gravitational field. The null results of this search set an upper limit on the coupling energy between the neutron spin and the gravity on the ground at 5.3$\times$10$^{-22}$~eV (95\% confidence level), resulting in a 17-fold improvement over the previous limit. The results can also be used to constrain several other anomalous interactions. In particular, the limit on the coupling strength of axion-mediated monopole-dipole interactions at the range of the Earth radius is improved by a factor of 17.
\end{abstract}

\maketitle

Among the four fundamental forces, the electromagnetic, the strong and the weak interactions are all dependent on particle spins according to the standard model of particle physics, only the gravitational interaction is spin independent according to the general theory of relativity. This principle should be tested experimentally with ever increasing precisions~\cite{ni2010}. At the same time, searches for spin-gravity coupling also test the fundamental symmetries of the gravitational interaction, since such coupling would break parity (P) and the time-reversal symmetry (T)~\cite{leitner1964,dass1976}. These symmetries are preserved in the electromagnetic and the strong interactions, but are broken in the weak interaction. Questions have been raised on the fundamental symmetry properties of the very weak gravitational interaction~\cite{leitner1964}.

The simplest form of spin-gravity coupling can be express as~\cite{leitner1964,Flambaum2009}:
\begin{equation}
\label{eq:Vsg}
V_{sg}(\boldsymbol{r})=\chi \boldsymbol{\sigma}\cdot {\boldsymbol{g}(\boldsymbol{r})},
\end{equation}
where $\chi$ is the coupling constant, $\hbar\mathbf{\sigma}$ is the particle spin, and ${\boldsymbol{g}(\boldsymbol{r})}$ is the gravitational acceleration at the location $\boldsymbol{r}$. This $P-$odd and $T-$odd coupling introduces a gravitational dipole moment to the particle, so that its center of mass and center of gravity are separated accordingly~\cite{peres1978}. Moreover, this coupling leads to a new force on the particle, $\boldsymbol{F}_a=-\nabla V_{sg}(\boldsymbol{r})$, which causes a spin-dependent local gravitational acceleration. In this way, the spin-gravity coupling violates the equivalence principle.

Spin-gravity coupling can also appear due to spin-mass coupling postulated in theories beyond the standard model.  Coupling between the 10$^{51}$ nucleons of the Earth and spins in the laboratory is a form of monopole-dipole interaction at the Earth range~\cite{Moody1984}. Such an interaction can be mediated by ultralight, axion-like, scalar bosons, which are candidates for cold dark matter in the Universe~\cite{ipser1983}. For two particles $a$ (monopole coupling, mass) and $b$ (dipole coupling, spin), the monopole-dipole interaction can be written as~\cite{Moody1984,Dobrescu2006,fadeev2019}:
\begin{equation}
\label{eq:Vmd}
V_{md}(r)=\frac{\hbar}{c}\frac{g_s^{a}g_p^{b}}{8 \pi m_b} \boldsymbol{\sigma}_b\cdot \hat{\boldsymbol{r}}\left(\frac{1}{r \lambda}+\frac{1}{r^{2}}\right)e^{-r / \lambda},
\end{equation}
where ${g_s}$ and ${g_p}$ are the scalar and pseudoscalar coupling strength, respectively. $m_b$ is the particle mass with the dipole coupling, $\hat{\boldsymbol{r}}$ is the unit vector connecting the two particles, and $\lambda$ is the reduced Compton wavelength of the interaction propagator.

Searching for spin-gravity coupling has motivated a wealth of experimental efforts. Tests of universal free fall of atoms with different nuclear spins~\cite{tarallo2014} or internal states~\cite{duan2016} were conducted with atom interferometers. Tests of local Lorentz invariance were performed using rotatable torsion balances and polarized massive objects~\cite{hou2001,Heckel2008}. Searches for atomic energy shifts correlated with the flipping of the quantization axis relative to the Earth gravity were also carried out, with $^9$Be$^+$ ions stored in a Penning trap~\cite{Wineland1991}, a $^{85}$Rb-$^{87}$Rb comagnetometer~\cite{kimball2017,kimball2023}, or a $^{199}$Hg-$^{201}$Hg comagnetometer~\cite{Venema1992}. The most stringent upper limits on the spin-gravity coupling strength of the neutron has been set by Ref.~\cite{Venema1992}.

Meanwhile, searching for monopole-dipole interactions are dominated by the use of atomic magnetometers and comagnetometers~\cite{safronova2018,terrano2021}. The most stringent laboratory limits on $\lvert g_s^Ng_p^n\rvert/\hbar c$ ($N$ and $n$ denote the nucleon and neutron, respectively) over the range of $\lambda >1$~m have been provided by experiments using a $^3$He-K self-compensation comagnetometer~\cite{Lee2018} and a $^{199}$Hg-$^{201}$Hg comagnetometer~\cite{Venema1992}. In addition, a model-dependent constraint has also been set by analyses of astronomical events~\cite{Raffelt2012}, which surpasses the laboratory limit when $\lambda < 2\times10^4$~m.

In this work, we employ a ground-based $^{129}$Xe-$^{131}$Xe-Rb atomic comagnetometer to search for the aforementioned exotic spin-dependent interactions between the neutron spin and the Earth. This comagnetometer configuration greatly suppresses the influence due to drifts and fluctuations in the bias field~\cite{terrano2021}. As a quantum compass, this comagnetometer is used to align the direction of the bias field along the Earth rotation axis with a precision of $\pm$ 0.58$^\circ$ (1$\sigma$), so that the systematic effect in setting the bias field orientation is minimized. By measuring the ratio of nuclear spin-precession frequencies between $^{129}$Xe and $^{131}$Xe as the bias field is flipped between being parallel and antiparallel to the Earth rotation direction, we determine the Earth rotation rate with an accuracy of $\pm$ 2.6 nHz (1$\sigma$). After subtracting off the Earth rotation effect, the remaining results lead to a limit improved by an order of magnitude on both the spin-dependent gravitational interaction and the monopole-dipole coupling for the neutron spin at the Earth range.

\begin{figure}
\includegraphics[width=3in]{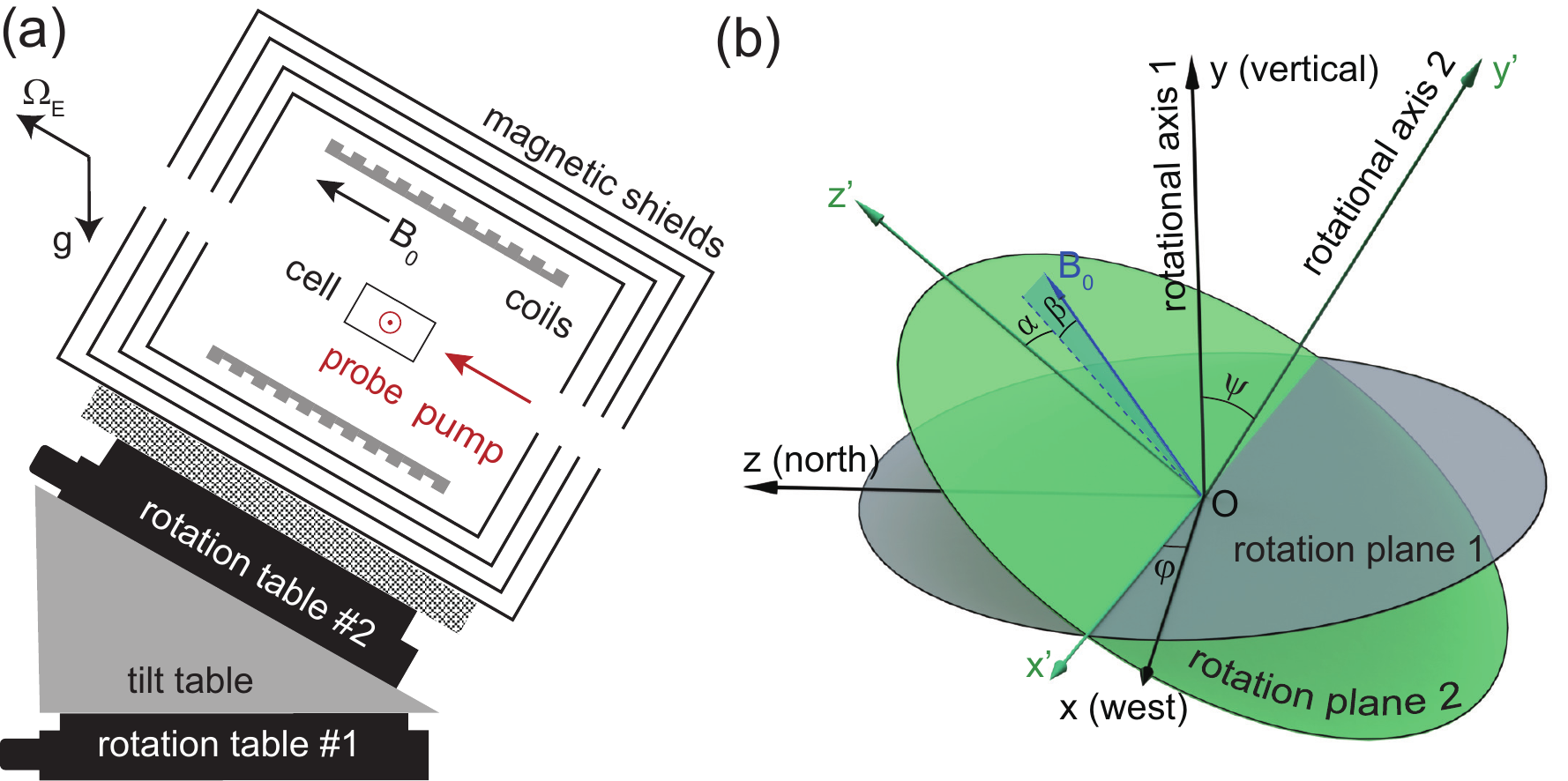}
\caption{\label{fig:setup} (a) Illustration of the experimental setup. The comagnetometer is mounted on a system of a horizontally placed rotation table (\#1), a tilt table, and a tilted rotation table (\#2).  (b) Coordinate systems of the setup. The $x$-$y$-$z$ system is defined by the geographical orientations, and corresponds to Table \#1; similarly, the $x’$-$y’$-$z’$ system corresponds to Table \#2. $\boldsymbol{\Omega}_E$ is contained in the $y$-$z$ plane by definition, and tilted by $\theta_L=31.82^\circ$ above the $z$ axis. $\mathbf{B}_0$ forms a small angle $\beta$ with the $x’$-$z’$ plane, and its projection on $x’$-$z’$ forms an angle $\alpha$ with $z’$. }
\end{figure}

The experiment is performed in Hefei, China, at the latitude of 31.82$^\circ$. The comagnetometer cell has an inner dimension of 10 $\times$ 8 $\times$ 8~mm$^3$. It is filled with Rb atoms of natural isotopic abundances, 4 Torr of $^{129}$Xe (nuclear spin $I$=1/2), 35 Torr of $^{131}$Xe ($I$=3/2), 5 Torr of H$_2$, and 160 Torr of N$_2$. The cell is placed at the center of a solenoid system, and four layers of mu-metal shields (Fig.~\ref{fig:setup}(a)). A uniform bias field $\mathbf{B}_0$ ($\sim$ 3.5~$\mu$T) is generated inside and pointing along the axis of the cylindrical shields. A circularly polarized ``pump” laser beam is directed along $\mathbf{B}_0$ to generate spin-polarized Rb atoms. A linearly polarized ``probe” laser beam is used to measure the Rb polarization component perpendicular to $\mathbf{B}_0$. Here, the polarized Rb atoms are used both to hyperpolarize the Xe atoms and to sense the nuclear spin signals of Xe atoms as an in-situ magnetometer ~\cite{sheng14a}. The two Xe isotopes are chosen for their long nuclear spin coherence times, and for their nearly equal collisional shifts in the Rb vapor~\cite{Feng2020}. To suppress the effect due to the nuclear quadrupole moment of $^{131}$Xe, we amplify and resolve the quadrupole splittings by deliberately employing the elongated cell geometry~\cite{Feng2020}. The entire comagnetometer system is mounted on a set of two rotation tables (\#1 and \#2) and a tilt table (Fig.~\ref{fig:setup}(a)).
%More details are given in Methods, Section I $-$ III.

The comagnetometer operates in a similar way as in Ref.~\cite{Feng2022}, and we extract the nuclear spin precession frequency $\omega$ by fitting the comagnetometer signals~\cite{supp}. There are three sources that contribute to $\boldsymbol{\omega}$: first, the Larmor precession, whose frequency $\boldsymbol{\omega}_L=\gamma\mathbf{B}_0$; second, the Earth rotation~\cite{walker16} at an angular velocity ${\boldsymbol{\Omega}_E}$; third, the anomalous spin-dependent couplings described in Eqs.~\eqref{eq:Vsg} and ~\eqref{eq:Vmd}.  It is customary to convert both  Eqs.~\eqref{eq:Vsg} and ~\eqref{eq:Vmd} into the simpler expression of $V=\epsilon\boldsymbol{I}\cdot\boldsymbol{A}$,
with $\epsilon$ as the fraction of the particle spin projected onto the atomic spin vector $\boldsymbol{I}$, and $\boldsymbol{A}$ pointing along the direction of the local gravity. In this way, the new physics can be treated as torque on Xe spins. The Larmor precession term is dominant so that $\omega$ can be approximately expressed as
\begin{equation}
\omega=\lvert\gamma{B}_0+\Omega_E\cos\theta+\textit{A}\epsilon \cos\phi\rvert,
\end{equation}
where $\theta$ is the angle between ${\boldsymbol{\Omega}_E}$  and ${\mathbf{B}}_0$, and $\phi$ is the angle between ${\boldsymbol{A}}$  and ${\mathbf{B}}_0$. The frequency ratio $R$ of the two Xe isotopes can be expressed as
\begin{equation}
~\label{eq:R}
R=\frac{\omega_{129}}{\omega_{131}}\approx-\rho-\frac{1-\rho}{\omega_{L,131}}\Omega_E\cos\theta-\frac{\epsilon_{129}-\rho\epsilon_{131}}{\omega_{L,131}}A\cos\phi,
\end{equation}
where $\omega_{L,131}$ is the Larmor precession frequency of $^{131}$Xe driven by its magnetic dipole moment. The ratio of gyromagnetic ratios $\rho=\gamma_{129}/\gamma_{131}=-3.37337(2)$ is determined in this experiment, and the result is consistent with that reported by Bulatowicz \textit{et al.}~\cite{Bulatowicz2013}. The Earth rotation frequency $\Omega_E/2\pi$=11605.761~nHz has been precisely determined with a sub-pHz error~\cite{groten2000}.

We aim to set the angle $\theta$ between the bias field ($\mathbf{B}_0$) and the Earth rotation direction ($\boldsymbol{\Omega}_E$) to be close to zero, where $\cos\theta$ is least sensitive to the angle-calibration uncertainty. Here we describe a compass procedure to determine the Earth rotation direction using the comagnetometer.

As shown in Eq.~(5) of Supplementary Materials~\cite{supp}, $\theta$ can be derived with the angles defined in Fig.~\ref{fig:setup}(b). In order to determine $\beta$, we send a circularly polarized calibration laser beam nearly parallel to $\mathbf{B}_0$ through the cell, and monitor its resonant absorption by Rb~\cite{supp}. Modulated absorption spectroscopy is performed to align $\mathbf{B}_0$ with the calibration laser beam. We then measure the angle between the laser beam and the $x’$-$z’$ plane of Table \#2. In this way, $\beta$ is calibrated to be $-0.14^\circ\pm0.30^\circ$. $\psi$ is then set to 31.96$^\circ$ $\pm$ 0.05$^\circ$ by adjusting the tilt table so that the central values of $\psi+\beta$ equals to the latitude angle $\theta_L$.

\begin{figure}[htb]
\centering
\includegraphics[width=3in]{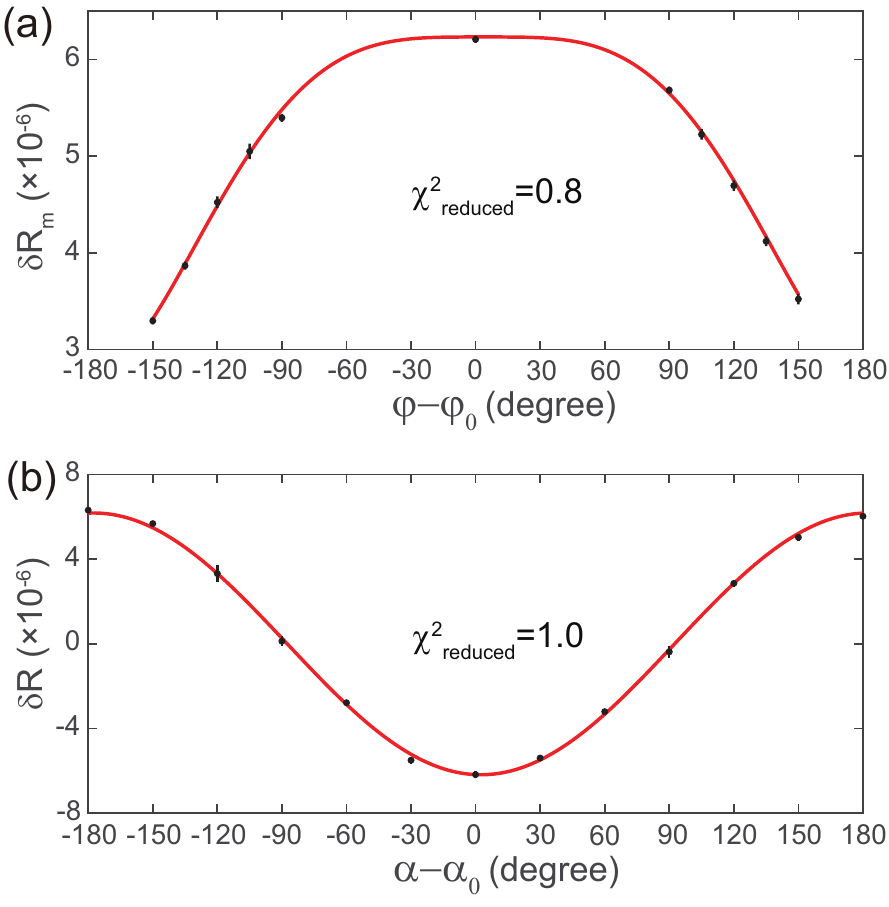}
\caption{\label{fig:calib} (a) Frequency-ratio amplitude $R_{amp}$ as a function of $\varphi$. For each data point, in order to determine the amplitude, $R$ is measured at 13 positions of $\alpha$, costing a measurement time of 26 hours. (b) Frequency ratio $R$ as a function of $\alpha$ with $\varphi$=0 and $\beta+\psi$=$\theta_L$. $R$ is shown as $\delta R(\alpha)=R(\alpha)-\overline{R(\alpha)}$. Each data point takes 2 hours to collect. The red lines in both plots are fitting results. $\varphi_0$ and $\alpha_0$ are offset angles. The data in both plots is taken with B$_0$=2.32~$\mu$T and $T=110^\circ$C. }
\end{figure}

The part of the measured frequency ratio $R$ that is dependent on $\alpha$ is
\begin{eqnarray}~\label{eq:Ra}
R_\alpha=&&-\frac{(1-\rho)\Omega_E}{\omega_{L,131}}\left\{\sin\theta_L\cos\alpha\sin\psi\cos\beta+\cos\theta_L\right.\nonumber\\
&&\left.[\cos\beta(\cos\alpha\cos\varphi\cos\psi-\sin\alpha\sin\varphi)]\right\},
\end{eqnarray}
which can also be expressed as $R_\alpha=R_{amp}(\varphi)\sin(\alpha-c)$ for each chosen $\varphi$ with $c$ as a phase offset. Fig.~\ref{fig:calib}(a) shows the experimental results of $R_{amp}$ at different rotation angles of Table \#1. By fitting the results using the relation based on Eq.~\eqref{eq:Ra} and repeating the experiment process three times, we determine the position of $\varphi=0^\circ$ with an uncertainty of $\pm$ 0.48$^\circ$ (1$\sigma$).

After $\varphi$ is fixed at zero, $R$ can be expressed as
\begin{equation}~\label{eq:Ralpha}
R=\lvert \rho\rvert-\frac{(1-\rho)\Omega_E}{\omega_{L,131}}(\sin^2\beta+\cos^2\beta\cos\alpha).
\end{equation}
By fitting the experimental results in Fig.~\ref{fig:calib}(b) using Eq.~\eqref{eq:Ralpha} and repeating the experiment process three times, we determine the position of $\alpha=0^\circ$ with an error of $\pm$ 0.28$^\circ$ (1$\sigma$). Combining all of the measurement uncertainties, we align the bias field $\mathbf{B}_0$ to the Earth rotation axis $\boldsymbol{\Omega}_E$ with the result of $\theta= 0^\circ \pm 0.58^\circ$ (1$\sigma$).

\begin{figure}[htb]
\centering
\includegraphics[width=3in]{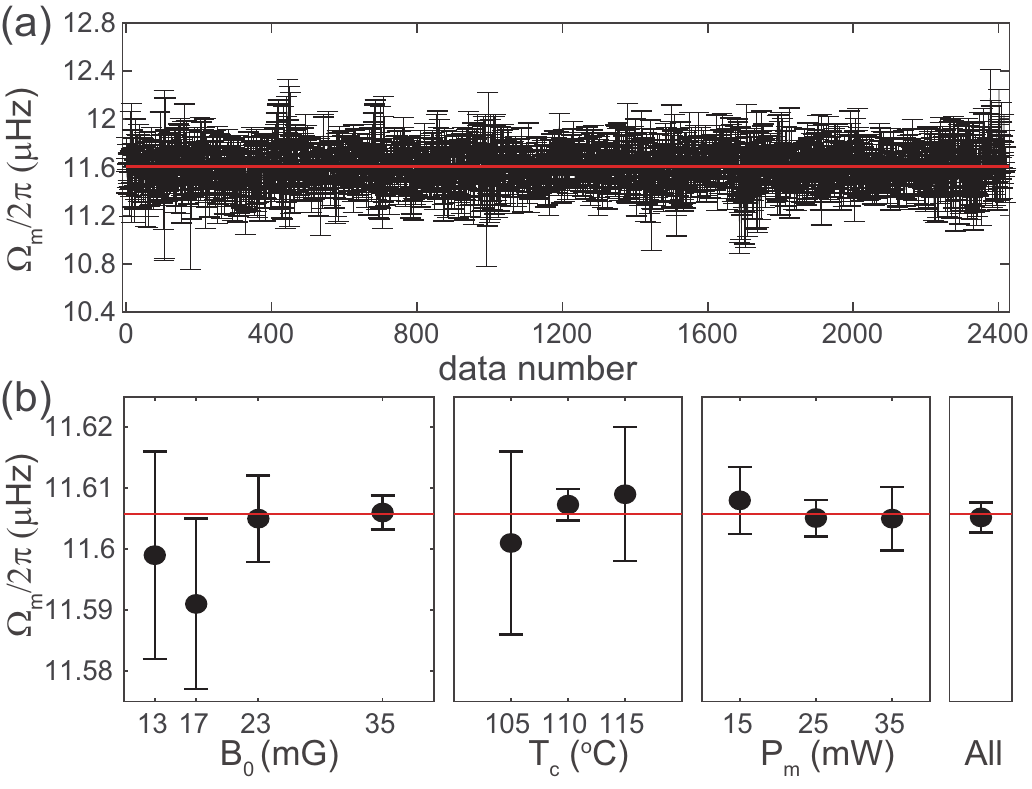}
\caption{\label{fig:longdata} (a) All the data collected in the search for anomalous couplings. Each data point represents an average result from 16 experiment cycles taken in over 60 minutes. The weighted average of all data is $\Omega_m/2\pi=11605.2 \pm 2.5$~(stat)~nHz, with the reduced $\chi^2$ as 1.2. (b) Studies of systematic effects by varying the bias field, cell temperature, pump beam powers, etc. The weighted average of all data is given in the window ``All". In both (a) and (b), the red lines mark the recommended value of the Earth rotation frequency.}
\end{figure}

Once the alignment is complete, the search for new physics starts by comparing the frequency ratios between $R_+$ at $\theta_+=0^\circ\pm 0.58^\circ$ and $R_-$ at $\theta_-=180^\circ + 2\beta$, and record their difference $\Delta R=R_--R_+$. We define the resolved rotation rate $\Omega_m$ of the comagnetometer as $\omega_{L,131}\Delta R/2(1-\rho)$ to describe the search results as this quantity is independent of the magnitude of $B_0$
\begin{eqnarray}
~\label{eq:dR}
\Omega_m=\frac{(\cos\theta_+-\cos\theta_-)\Omega_E}{2}+\frac{(\epsilon_{129}-\rho\epsilon_{131})A\cos\phi}{1-\rho}.
\end{eqnarray}
For each data point of $\Omega_m$ in Fig.~\ref{fig:longdata}(a), we spend four experimental cycles at $\theta_+$, followed by eight cycles at $\theta_-$, then again with four cycles at $\theta_+$, for a total of 16 cycles over a total of 60 minutes including the time spent on the rotation of Table \#2. Our comagnetometer achieves a typical rotation sensitivity of $1\times 10^{-7}$ Hz $\cdot$ hr$^{1/2}$ on $\Omega_m/2\pi$.

$\Omega_E$ is precisely known, but the measured values of $\theta$ and $\beta$ lead to a correction and error of -0.68 $\pm$ 0.52~nHz on the $\Omega_E$ term on the right hand side of Eq.~\eqref{eq:dR}. While this is the dominant systematic effect to determine the new physics, other sources of systematic uncertainties are also investigated~\cite{supp}. We deliberately vary experimental parameters, including $B_0$, the oven temperature, the pump beam power, and the $\pi/2$ pulse amplitude, and find no effects on the measured $\Omega_m$ values.

We plot all the data taken over a span of three months in Fig.~\ref{fig:longdata}(a), and selectively plot several systematic studies in  Fig.~\ref{fig:longdata}(b). The weighted average of all the data is $\Omega_m/2\pi$=11605.0 $\pm$ 2.5(stat) $\pm$ 0.2(sys)~nHz, leading to an independent measurement of the Earth rotation rate $\Omega_{E,m}/2\pi$=11605.7 $\pm$ 2.5(stat) $\pm$ 0.6(sys)~nHz. Though the precision of $3\times10^{-4}$ achieved in this work is far lower than that of the state-of-the-art Very Large Ring Laser Gyroscope~\cite{schreiber2013}, it is one order of magnitude better than the previous best results using either comagnetometers~\cite{Venema1992,gemmel2010} or atomic interferometers~\cite{dickerson2013,yao2018}.

As in Ref.~\cite{Venema1992}, we use the Schmidt model~\cite{Schmidt1937,kimball2015} for nuclear spin analysis, and get $\epsilon_{129}=+1$ and $\epsilon_{131}=-0.2$. Combining these results and Eq.~\eqref{eq:dR}, we extract the spin-gravity coupling parameter $\lvert A_{sg}\rvert/2\pi=3.1 \pm 65.4$~nHz, with the error budget of $A_{sg}$ listed in Table~\ref{tab:errA}. The energy difference between a spin-up and a spin-down state of a neutron on the ground, $\hbar\lvert A_{sg}\rvert$, is less than  $5.3\times10^{-22}$~eV at the 95\% confidence level (C.L.). As shown in Table~\ref{tab:sg}, this limit on the spin-gravity coupling of the neutron is improved over the previous best limit by a factor of 17~\cite{Venema1992,venemathesis}. This result also leads to an upper limit of 2.7 fm (95\% C.L.) on the separation ($\hbar\lvert A_{sg}\rvert/2mg$) between the center of mass and the center of gravity of the neutron.

\begin{table}[htb]
\begin{center}
\caption{\label{tab:errA} Error budget of the spin-gravity coupling parameter $A_{sg}/2\pi$.}
%\begin{ruledtabular}
\begin{tabular}{lccc}
\hline
 &Correction & Uncertainty \\
&(nHz)&(nHz)\\
\hline
Bias field alignment  & 17.3& 13.3\\
Cell temperature correlation  & -3.8    & 5.1\\
Residual magnetic field  & -1.3  & 1.4\\
Pump beam power correlation &$<$0.1  & $<$0.1\\
\hline
Systematics total & 12.2 & 14.3\\
Statistical result & & 63.8\\
\hline
Total  &12.2&65.4\\
\hline
\end{tabular}
%\end{ruledtabular}
\end{center}
\end{table}

\begin{table}
\caption{\label{tab:sg} Constraints (95\% C.L.) on the energy difference $\hbar\lvert{A_{sg}}\rvert$ due to the spin-gravity coupling  and the E$\ddot{\mathrm{o}}$tv$\ddot{\mathrm{o}}$s parameter $\eta_s$.}
%\begin{ruledtabular}
\begin{tabular}{ccccc}
\hline
System & Spin &$\hbar\lvert{A_{sg}}\rvert$~(eV)& $\lvert\eta_s\rvert$ & Reference \\
\hline
AlNiCo-SmCo$_5$&Electron & $2.2\times10^{-19}$&1.2$\times10^{-15}$&Ref.~\cite{Heckel2008}\\
$^{85}$Rb-$^{87}$Rb&Proton & $3.4\times10^{-18}$&1.1$\times10^{-17}$&Ref.~\cite{kimball2017}\\
\hline
$^9$Be$^+$&Neutron & $1.7\times10^{-19}$&5.4$\times10^{-19}$&Ref.~\cite{Wineland1991}\\
$^{199}$Hg-$^{201}$Hg&Neutron& $9.1\times10^{-21}$&2.9$\times10^{-20}$&Ref.~\cite{Venema1992}\\
$^{129}$Xe-$^{131}$Xe&Neutron&$5.3\times10^{-22}$&1.7$\times10^{-21}$&This work\\
%&$\sigma_{Xe}\cdot\hat{g}$&$4.8\times10^{-23}$\\
\hline
%Planck scale~\cite{Flambaum2009} & $h|A_{sg}|$ &$4.3\times10^{-23}$\\
%Planck-scale coupling\\
\end{tabular}
\end{table}

The spin-gravity coupling also leads to a spin-dependent term in the gravitational acceleration.  For the two spin states of the neutron, the difference in the acceleration on the ground is $\lvert\delta g_s(r_E)\rvert=2\hbar\lvert A_{sg}\rvert/m_nr_E$, with $m_n$ as the neutron mass and $r_E$ as the Earth radius. The E$\ddot{\mathrm{o}}$tv$\ddot{\mathrm{o}}$s parameter of the neutron is defined as
\begin{equation}
\lvert\eta_{s,n}\rvert=\frac{\lvert\delta g_{s,n}(r_E)\rvert}{g(r_E)}=\frac{2\hbar\lvert A_{sg,n}(r_E)\rvert}{m_ng(r_E)r_E}.
\end{equation}
The results of this work provide an upper limit $\lvert\eta_{s,n}\rvert\leq1.7\times10^{-21}$ (95\% C.L.). In comparison, free-fall experiments with atom interferometers have placed limits of $\lvert\eta_{s}\rvert$ only at the 10$^{-7}$ level~\cite{tarallo2014,duan2016}, although there the accelerations were measured more directly.

\begin{figure}[htb]
\centering
\includegraphics[width=3.0in]{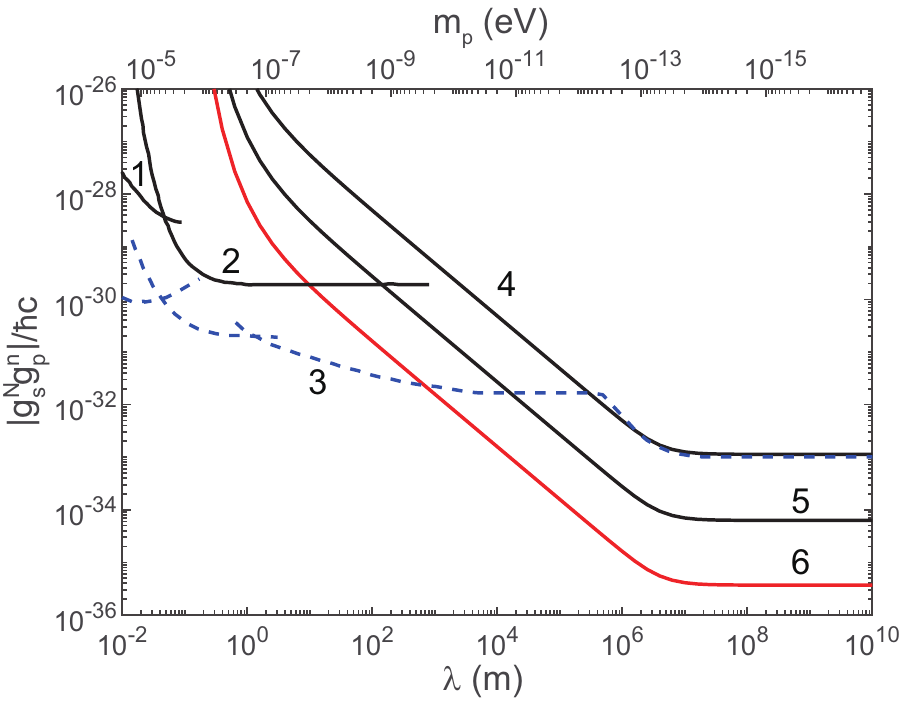}
\caption{\label{fig:g} The upper limits (95\% C.L.) on the monopole-dipole coupling constants $\lvert g_s^Ng_p^n\rvert/\hbar c$. Line 1 is based on a $^3$He-$^{129}$Xe comagnetometer~\cite{Tullney2013}, line 2 uses a self-compensating $^3$He-K comagnetometer~\cite{Lee2018}, line 3 is from the analysis of astronomical observation~\cite{Raffelt2012,miscRaffelt}, line 4 is from  the spectroscopy of trapped Be$^{+}$ ions~\cite{Wineland1991}, line 5 is from a $^{199}$Hg-$^{201}$Hg comagnetometer~\cite{Venema1992}, and line 6 is from this work.}
\end{figure}

For the monopole-dipole interaction in Eq.~\eqref{eq:Vmd}, the constraint set by this work on the coupling constants $\lvert g_s^Ng_p^n\rvert/\hbar c$ surpasses the results extracted from the astronomical events at $\lambda>1\times10^3$~m (propagator mass $m_p<2.0\times10^{-10}$~eV) (Fig.~\ref{fig:g}), and reaches $3.7\times10^{-36}$ (95\% C.L.) when $\lambda>1\times10^8$~m ($m_p<2.0\times10^{-15}$~eV), which is a 17-fold improvement over the previous best limit~\cite{Venema1992}.

The results of this experiment can also be used to study other related theoretical models of anomalous interactions~\cite{safronova2018,krause2023}. One class of these models treat the Earth as a source of polarized electrons~\cite{Hunter2013,poddar2023}. These electrons exist in iron-containing minerals, and align along the Earth
magnetic field lines~\cite{Hunter2013}. Therefore, the results of this work can be used to search for a possible coupling between the electron spin and the neutron spin mediated by spin-1 vector bosons~\cite{Almasi2020}.  In addition to the improved measurements realized in this work, the spatial distribution of the polarized geo-electrons under Hefei needs to be modelled before a new constraint on this coupling can be reached~\cite{zhangprep}.

The search sensitivity can be further improved by: installing a multipass cavity~\cite{hao2021} for the probe laser beam to improve the signal-to-noise ratio, implementing an integrated hardware design for better mechanical stability, reducing the uncertainty in the orientation of the bias field with the help of external references such as a fiber laser gyroscope, and accumulating more data with a longer measurement time.

A part of this work is carried out at the USTC Center for Micro and Nanoscale Research and Fabrication. We thank Professor D. F. J. Kimball, Professor L. Hunter, Dr. M. Limes, Professor L. A. Orozco, and Professor M. V. Romalis for helpful discussions.  This work was supported by Natural Science Foundation of China (Grant No. 12174372) and the Strategic Priority Research Program, CAS (No. XDB21010200).


\begin{thebibliography}{46}%
\makeatletter
\providecommand \@ifxundefined [1]{%
 \@ifx{#1\undefined}
}%
\providecommand \@ifnum [1]{%
 \ifnum #1\expandafter \@firstoftwo
 \else \expandafter \@secondoftwo
 \fi
}%
\providecommand \@ifx [1]{%
 \ifx #1\expandafter \@firstoftwo
 \else \expandafter \@secondoftwo
 \fi
}%
\providecommand \natexlab [1]{#1}%
\providecommand \enquote  [1]{``#1''}%
\providecommand \bibnamefont  [1]{#1}%
\providecommand \bibfnamefont [1]{#1}%
\providecommand \citenamefont [1]{#1}%
\providecommand \href@noop [0]{\@secondoftwo}%
\providecommand \href [0]{\begingroup \@sanitize@url \@href}%
\providecommand \@href[1]{\@@startlink{#1}\@@href}%
\providecommand \@@href[1]{\endgroup#1\@@endlink}%
\providecommand \@sanitize@url [0]{\catcode `\\12\catcode `\$12\catcode
  `\&12\catcode `\#12\catcode `\^12\catcode `\_12\catcode `\%12\relax}%
\providecommand \@@startlink[1]{}%
\providecommand \@@endlink[0]{}%
\providecommand \url  [0]{\begingroup\@sanitize@url \@url }%
\providecommand \@url [1]{\endgroup\@href {#1}{\urlprefix }}%
\providecommand \urlprefix  [0]{URL }%
\providecommand \Eprint [0]{\href }%
\providecommand \doibase [0]{https://doi.org/}%
\providecommand \selectlanguage [0]{\@gobble}%
\providecommand \bibinfo  [0]{\@secondoftwo}%
\providecommand \bibfield  [0]{\@secondoftwo}%
\providecommand \translation [1]{[#1]}%
\providecommand \BibitemOpen [0]{}%
\providecommand \bibitemStop [0]{}%
\providecommand \bibitemNoStop [0]{.\EOS\space}%
\providecommand \EOS [0]{\spacefactor3000\relax}%
\providecommand \BibitemShut  [1]{\csname bibitem#1\endcsname}%
\let\auto@bib@innerbib\@empty
%</preamble>
\bibitem [{\citenamefont {Ni}(2010)}]{ni2010}%
  \BibitemOpen
  \bibfield  {author} {\bibinfo {author} {\bibfnamefont {W.-T.}\ \bibnamefont
  {Ni}},\ }\bibfield  {title} {\bibinfo {title} {Searches for the role of spin
  and polarization in gravity},\ }\href
  {https://doi.org/10.1088/0034-4885/73/5/056901} {\bibfield  {journal}
  {\bibinfo  {journal} {Reports on Progress in Physics}\ }\textbf {\bibinfo
  {volume} {73}},\ \bibinfo {pages} {056901} (\bibinfo {year}
  {2010})}\BibitemShut {NoStop}%
\bibitem [{\citenamefont {Leitner}\ and\ \citenamefont
  {Okubo}(1964)}]{leitner1964}%
  \BibitemOpen
  \bibfield  {author} {\bibinfo {author} {\bibfnamefont {J.}~\bibnamefont
  {Leitner}}\ and\ \bibinfo {author} {\bibfnamefont {S.}~\bibnamefont
  {Okubo}},\ }\bibfield  {title} {\bibinfo {title} {Parity, charge conjugation,
  and time reversal in the gravitational interaction},\ }\href
  {https://doi.org/10.1103/PhysRev.136.B1542} {\bibfield  {journal} {\bibinfo
  {journal} {Phys. Rev.}\ }\textbf {\bibinfo {volume} {136}},\ \bibinfo {pages}
  {B1542} (\bibinfo {year} {1964})}\BibitemShut {NoStop}%
\bibitem [{\citenamefont {Dass}(1976)}]{dass1976}%
  \BibitemOpen
  \bibfield  {author} {\bibinfo {author} {\bibfnamefont {N.~D.~H.}\
  \bibnamefont {Dass}},\ }\bibfield  {title} {\bibinfo {title} {Test for {$C$},
  {$P$}, and {$T$} nonconservation in gravitation},\ }\href
  {https://doi.org/10.1103/PhysRevLett.36.393} {\bibfield  {journal} {\bibinfo
  {journal} {Phys. Rev. Lett.}\ }\textbf {\bibinfo {volume} {36}},\ \bibinfo
  {pages} {393} (\bibinfo {year} {1976})}\BibitemShut {NoStop}%
\bibitem [{\citenamefont {Flambaum}\ \emph {et~al.}(2009)\citenamefont
  {Flambaum}, \citenamefont {Lambert},\ and\ \citenamefont
  {Pospelov}}]{Flambaum2009}%
  \BibitemOpen
  \bibfield  {author} {\bibinfo {author} {\bibfnamefont {V.}~\bibnamefont
  {Flambaum}}, \bibinfo {author} {\bibfnamefont {S.}~\bibnamefont {Lambert}},\
  and\ \bibinfo {author} {\bibfnamefont {M.}~\bibnamefont {Pospelov}},\
  }\bibfield  {title} {\bibinfo {title} {Scalar-tensor theories with
  pseudoscalar couplings},\ }\href {https://doi.org/10.1103/PhysRevD.80.105021}
  {\bibfield  {journal} {\bibinfo  {journal} {Phys. Rev. D}\ }\textbf {\bibinfo
  {volume} {80}},\ \bibinfo {pages} {105021} (\bibinfo {year}
  {2009})}\BibitemShut {NoStop}%
\bibitem [{\citenamefont {Peres}(1978)}]{peres1978}%
  \BibitemOpen
  \bibfield  {author} {\bibinfo {author} {\bibfnamefont {A.}~\bibnamefont
  {Peres}},\ }\bibfield  {title} {\bibinfo {title} {Test of equivalence
  principle for particles with spin},\ }\href
  {https://doi.org/10.1103/PhysRevD.18.2739} {\bibfield  {journal} {\bibinfo
  {journal} {Phys. Rev. D}\ }\textbf {\bibinfo {volume} {18}},\ \bibinfo
  {pages} {2739} (\bibinfo {year} {1978})}\BibitemShut {NoStop}%
\bibitem [{\citenamefont {Moody}\ and\ \citenamefont
  {Wilczek}(1984)}]{Moody1984}%
  \BibitemOpen
  \bibfield  {author} {\bibinfo {author} {\bibfnamefont {J.~E.}\ \bibnamefont
  {Moody}}\ and\ \bibinfo {author} {\bibfnamefont {F.}~\bibnamefont
  {Wilczek}},\ }\bibfield  {title} {\bibinfo {title} {New macroscopic
  forces?},\ }\href {https://doi.org/10.1103/PhysRevD.30.130} {\bibfield
  {journal} {\bibinfo  {journal} {Phys. Rev. D}\ }\textbf {\bibinfo {volume}
  {30}},\ \bibinfo {pages} {130} (\bibinfo {year} {1984})}\BibitemShut
  {NoStop}%
\bibitem [{\citenamefont {Ipser}\ and\ \citenamefont
  {Sikivie}(1983)}]{ipser1983}%
  \BibitemOpen
  \bibfield  {author} {\bibinfo {author} {\bibfnamefont {J.}~\bibnamefont
  {Ipser}}\ and\ \bibinfo {author} {\bibfnamefont {P.}~\bibnamefont
  {Sikivie}},\ }\bibfield  {title} {\bibinfo {title} {Can galactic halos be
  made of axions?},\ }\href {https://doi.org/10.1103/PhysRevLett.50.925}
  {\bibfield  {journal} {\bibinfo  {journal} {Phys. Rev. Lett.}\ }\textbf
  {\bibinfo {volume} {50}},\ \bibinfo {pages} {925} (\bibinfo {year}
  {1983})}\BibitemShut {NoStop}%
\bibitem [{\citenamefont {Dobrescu}\ and\ \citenamefont
  {Mocioiu}(2006)}]{Dobrescu2006}%
  \BibitemOpen
  \bibfield  {author} {\bibinfo {author} {\bibfnamefont {B.~A.}\ \bibnamefont
  {Dobrescu}}\ and\ \bibinfo {author} {\bibfnamefont {I.}~\bibnamefont
  {Mocioiu}},\ }\bibfield  {title} {\bibinfo {title} {Spin-dependent
  macroscopic forces from new particle exchange},\ }\href
  {https://doi.org/10.1088/1126-6708/2006/11/005} {\bibfield  {journal}
  {\bibinfo  {journal} {Journal of High Energy Physics}\ }\textbf {\bibinfo
  {volume} {2006}},\ \bibinfo {pages} {005} (\bibinfo {year}
  {2006})}\BibitemShut {NoStop}%
\bibitem [{\citenamefont {Fadeev}\ \emph {et~al.}(2019)\citenamefont {Fadeev},
  \citenamefont {Stadnik}, \citenamefont {Ficek}, \citenamefont {Kozlov},
  \citenamefont {Flambaum},\ and\ \citenamefont {Budker}}]{fadeev2019}%
  \BibitemOpen
  \bibfield  {author} {\bibinfo {author} {\bibfnamefont {P.}~\bibnamefont
  {Fadeev}}, \bibinfo {author} {\bibfnamefont {Y.~V.}\ \bibnamefont {Stadnik}},
  \bibinfo {author} {\bibfnamefont {F.}~\bibnamefont {Ficek}}, \bibinfo
  {author} {\bibfnamefont {M.~G.}\ \bibnamefont {Kozlov}}, \bibinfo {author}
  {\bibfnamefont {V.~V.}\ \bibnamefont {Flambaum}},\ and\ \bibinfo {author}
  {\bibfnamefont {D.}~\bibnamefont {Budker}},\ }\bibfield  {title} {\bibinfo
  {title} {Revisiting spin-dependent forces mediated by new bosons: Potentials
  in the coordinate-space representation for macroscopic- and atomic-scale
  experiments},\ }\href {https://doi.org/10.1103/PhysRevA.99.022113} {\bibfield
   {journal} {\bibinfo  {journal} {Phys. Rev. A}\ }\textbf {\bibinfo {volume}
  {99}},\ \bibinfo {pages} {022113} (\bibinfo {year} {2019})}\BibitemShut
  {NoStop}%
\bibitem [{\citenamefont {Tarallo}\ \emph {et~al.}(2014)\citenamefont
  {Tarallo}, \citenamefont {Mazzoni}, \citenamefont {Poli}, \citenamefont
  {Sutyrin}, \citenamefont {Zhang},\ and\ \citenamefont {Tino}}]{tarallo2014}%
  \BibitemOpen
  \bibfield  {author} {\bibinfo {author} {\bibfnamefont {M.~G.}\ \bibnamefont
  {Tarallo}}, \bibinfo {author} {\bibfnamefont {T.}~\bibnamefont {Mazzoni}},
  \bibinfo {author} {\bibfnamefont {N.}~\bibnamefont {Poli}}, \bibinfo {author}
  {\bibfnamefont {D.~V.}\ \bibnamefont {Sutyrin}}, \bibinfo {author}
  {\bibfnamefont {X.}~\bibnamefont {Zhang}},\ and\ \bibinfo {author}
  {\bibfnamefont {G.~M.}\ \bibnamefont {Tino}},\ }\bibfield  {title} {\bibinfo
  {title} {Test of einstein equivalence principle for 0-spin and
  half-integer-spin atoms: Search for spin-gravity coupling effects},\ }\href
  {https://doi.org/10.1103/PhysRevLett.113.023005} {\bibfield  {journal}
  {\bibinfo  {journal} {Phys. Rev. Lett.}\ }\textbf {\bibinfo {volume} {113}},\
  \bibinfo {pages} {023005} (\bibinfo {year} {2014})}\BibitemShut {NoStop}%
\bibitem [{\citenamefont {Duan}\ \emph {et~al.}(2016)\citenamefont {Duan},
  \citenamefont {Deng}, \citenamefont {Zhou}, \citenamefont {Zhang},
  \citenamefont {Xu}, \citenamefont {Xiong}, \citenamefont {Xu}, \citenamefont
  {Shao}, \citenamefont {Luo},\ and\ \citenamefont {Hu}}]{duan2016}%
  \BibitemOpen
  \bibfield  {author} {\bibinfo {author} {\bibfnamefont {X.-C.}\ \bibnamefont
  {Duan}}, \bibinfo {author} {\bibfnamefont {X.-B.}\ \bibnamefont {Deng}},
  \bibinfo {author} {\bibfnamefont {M.-K.}\ \bibnamefont {Zhou}}, \bibinfo
  {author} {\bibfnamefont {K.}~\bibnamefont {Zhang}}, \bibinfo {author}
  {\bibfnamefont {W.-J.}\ \bibnamefont {Xu}}, \bibinfo {author} {\bibfnamefont
  {F.}~\bibnamefont {Xiong}}, \bibinfo {author} {\bibfnamefont {Y.-Y.}\
  \bibnamefont {Xu}}, \bibinfo {author} {\bibfnamefont {C.-G.}\ \bibnamefont
  {Shao}}, \bibinfo {author} {\bibfnamefont {J.}~\bibnamefont {Luo}},\ and\
  \bibinfo {author} {\bibfnamefont {Z.-K.}\ \bibnamefont {Hu}},\ }\bibfield
  {title} {\bibinfo {title} {Test of the universality of free fall with atoms
  in different spin orientations},\ }\href
  {https://doi.org/10.1103/PhysRevLett.117.023001} {\bibfield  {journal}
  {\bibinfo  {journal} {Phys. Rev. Lett.}\ }\textbf {\bibinfo {volume} {117}},\
  \bibinfo {pages} {023001} (\bibinfo {year} {2016})}\BibitemShut {NoStop}%
\bibitem [{\citenamefont {Hou}\ and\ \citenamefont {Ni}(2001)}]{hou2001}%
  \BibitemOpen
  \bibfield  {author} {\bibinfo {author} {\bibfnamefont {L.-S.}\ \bibnamefont
  {Hou}}\ and\ \bibinfo {author} {\bibfnamefont {W.-T.}\ \bibnamefont {Ni}},\
  }\bibfield  {title} {\bibinfo {title} {Rotatable-torsion-balance equivalence
  principle experiment for the spin-polarized {HoFe$_3$}},\ }\href
  {https://doi.org/10.1142/S0217732301003619} {\bibfield  {journal} {\bibinfo
  {journal} {Modern Physics Letters A}\ }\textbf {\bibinfo {volume} {16}},\
  \bibinfo {pages} {763} (\bibinfo {year} {2001})}\BibitemShut {NoStop}%
\bibitem [{\citenamefont {Heckel}\ \emph {et~al.}(2008)\citenamefont {Heckel},
  \citenamefont {Adelberger}, \citenamefont {Cramer}, \citenamefont {Cook},
  \citenamefont {Schlamminger},\ and\ \citenamefont {Schmidt}}]{Heckel2008}%
  \BibitemOpen
  \bibfield  {author} {\bibinfo {author} {\bibfnamefont {B.~R.}\ \bibnamefont
  {Heckel}}, \bibinfo {author} {\bibfnamefont {E.~G.}\ \bibnamefont
  {Adelberger}}, \bibinfo {author} {\bibfnamefont {C.~E.}\ \bibnamefont
  {Cramer}}, \bibinfo {author} {\bibfnamefont {T.~S.}\ \bibnamefont {Cook}},
  \bibinfo {author} {\bibfnamefont {S.}~\bibnamefont {Schlamminger}},\ and\
  \bibinfo {author} {\bibfnamefont {U.}~\bibnamefont {Schmidt}},\ }\bibfield
  {title} {\bibinfo {title} {Preferred-frame and {$CP$}-violation tests with
  polarized electrons},\ }\href {https://doi.org/10.1103/PhysRevD.78.092006}
  {\bibfield  {journal} {\bibinfo  {journal} {Phys. Rev. D}\ }\textbf {\bibinfo
  {volume} {78}},\ \bibinfo {pages} {092006} (\bibinfo {year}
  {2008})}\BibitemShut {NoStop}%
\bibitem [{\citenamefont {Wineland}\ \emph {et~al.}(1991)\citenamefont
  {Wineland}, \citenamefont {Bollinger}, \citenamefont {Heinzen}, \citenamefont
  {Itano},\ and\ \citenamefont {Raizen}}]{Wineland1991}%
  \BibitemOpen
  \bibfield  {author} {\bibinfo {author} {\bibfnamefont {D.~J.}\ \bibnamefont
  {Wineland}}, \bibinfo {author} {\bibfnamefont {J.~J.}\ \bibnamefont
  {Bollinger}}, \bibinfo {author} {\bibfnamefont {D.~J.}\ \bibnamefont
  {Heinzen}}, \bibinfo {author} {\bibfnamefont {W.~M.}\ \bibnamefont {Itano}},\
  and\ \bibinfo {author} {\bibfnamefont {M.~G.}\ \bibnamefont {Raizen}},\
  }\bibfield  {title} {\bibinfo {title} {Search for anomalous spin-dependent
  forces using stored-ion spectroscopy},\ }\href
  {https://doi.org/10.1103/PhysRevLett.67.1735} {\bibfield  {journal} {\bibinfo
   {journal} {Phys. Rev. Lett.}\ }\textbf {\bibinfo {volume} {67}},\ \bibinfo
  {pages} {1735} (\bibinfo {year} {1991})}\BibitemShut {NoStop}%
\bibitem [{\citenamefont {Jackson~Kimball}\ \emph {et~al.}(2017)\citenamefont
  {Jackson~Kimball}, \citenamefont {Dudley}, \citenamefont {Li}, \citenamefont
  {Patel},\ and\ \citenamefont {Valdez}}]{kimball2017}%
  \BibitemOpen
  \bibfield  {author} {\bibinfo {author} {\bibfnamefont {D.~F.}\ \bibnamefont
  {Jackson~Kimball}}, \bibinfo {author} {\bibfnamefont {J.}~\bibnamefont
  {Dudley}}, \bibinfo {author} {\bibfnamefont {Y.}~\bibnamefont {Li}}, \bibinfo
  {author} {\bibfnamefont {D.}~\bibnamefont {Patel}},\ and\ \bibinfo {author}
  {\bibfnamefont {J.}~\bibnamefont {Valdez}},\ }\bibfield  {title} {\bibinfo
  {title} {Constraints on long-range spin-gravity and monopole-dipole couplings
  of the proton},\ }\href {https://doi.org/10.1103/PhysRevD.96.075004}
  {\bibfield  {journal} {\bibinfo  {journal} {Phys. Rev. D}\ }\textbf {\bibinfo
  {volume} {96}},\ \bibinfo {pages} {075004} (\bibinfo {year}
  {2017})}\BibitemShut {NoStop}%
\bibitem [{\citenamefont {Jackson~Kimball}\ \emph {et~al.}(2023)\citenamefont
  {Jackson~Kimball}, \citenamefont {Dudley}, \citenamefont {Li}, \citenamefont
  {Patel},\ and\ \citenamefont {Valdez}}]{kimball2023}%
  \BibitemOpen
  \bibfield  {author} {\bibinfo {author} {\bibfnamefont {D.~F.}\ \bibnamefont
  {Jackson~Kimball}}, \bibinfo {author} {\bibfnamefont {J.}~\bibnamefont
  {Dudley}}, \bibinfo {author} {\bibfnamefont {Y.}~\bibnamefont {Li}}, \bibinfo
  {author} {\bibfnamefont {D.}~\bibnamefont {Patel}},\ and\ \bibinfo {author}
  {\bibfnamefont {J.}~\bibnamefont {Valdez}},\ }\bibfield  {title} {\bibinfo
  {title} {Erratum: Constraints on long-range spin-gravity and monopole-dipole
  couplings of the proton [{Phys.} {Rev.} {D} 96, 075004 (2017)]},\ }\href
  {https://doi.org/10.1103/PhysRevD.107.019903} {\bibfield  {journal} {\bibinfo
   {journal} {Phys. Rev. D}\ }\textbf {\bibinfo {volume} {107}},\ \bibinfo
  {pages} {019903} (\bibinfo {year} {2023})}\BibitemShut {NoStop}%
\bibitem [{\citenamefont {Venema}\ \emph {et~al.}(1992)\citenamefont {Venema},
  \citenamefont {Majumder}, \citenamefont {Lamoreaux}, \citenamefont {Heckel},\
  and\ \citenamefont {Fortson}}]{Venema1992}%
  \BibitemOpen
  \bibfield  {author} {\bibinfo {author} {\bibfnamefont {B.~J.}\ \bibnamefont
  {Venema}}, \bibinfo {author} {\bibfnamefont {P.~K.}\ \bibnamefont
  {Majumder}}, \bibinfo {author} {\bibfnamefont {S.~K.}\ \bibnamefont
  {Lamoreaux}}, \bibinfo {author} {\bibfnamefont {B.~R.}\ \bibnamefont
  {Heckel}},\ and\ \bibinfo {author} {\bibfnamefont {E.~N.}\ \bibnamefont
  {Fortson}},\ }\bibfield  {title} {\bibinfo {title} {Search for a coupling of
  the earths gravitational-field to nuclear spins in atomic mercury},\ }\href
  {https://doi.org/DOI 10.1103/PhysRevLett.68.135} {\bibfield  {journal}
  {\bibinfo  {journal} {Phys. Rev. Lett.}\ }\textbf {\bibinfo {volume} {68}},\
  \bibinfo {pages} {135} (\bibinfo {year} {1992})}\BibitemShut {NoStop}%
\bibitem [{\citenamefont {Safronova}\ \emph {et~al.}(2018)\citenamefont
  {Safronova}, \citenamefont {Budker}, \citenamefont {DeMille}, \citenamefont
  {Kimball}, \citenamefont {Derevianko},\ and\ \citenamefont
  {Clark}}]{safronova2018}%
  \BibitemOpen
  \bibfield  {author} {\bibinfo {author} {\bibfnamefont {M.~S.}\ \bibnamefont
  {Safronova}}, \bibinfo {author} {\bibfnamefont {D.}~\bibnamefont {Budker}},
  \bibinfo {author} {\bibfnamefont {D.}~\bibnamefont {DeMille}}, \bibinfo
  {author} {\bibfnamefont {D.~F.~J.}\ \bibnamefont {Kimball}}, \bibinfo
  {author} {\bibfnamefont {A.}~\bibnamefont {Derevianko}},\ and\ \bibinfo
  {author} {\bibfnamefont {C.~W.}\ \bibnamefont {Clark}},\ }\bibfield  {title}
  {\bibinfo {title} {Search for new physics with atoms and molecules},\ }\href
  {https://doi.org/10.1103/RevModPhys.90.025008} {\bibfield  {journal}
  {\bibinfo  {journal} {Rev. Mod. Phys.}\ }\textbf {\bibinfo {volume} {90}},\
  \bibinfo {pages} {025008} (\bibinfo {year} {2018})}\BibitemShut {NoStop}%
\bibitem [{\citenamefont {Terrano}\ and\ \citenamefont
  {Romalis}(2021)}]{terrano2021}%
  \BibitemOpen
  \bibfield  {author} {\bibinfo {author} {\bibfnamefont {W.~A.}\ \bibnamefont
  {Terrano}}\ and\ \bibinfo {author} {\bibfnamefont {M.~V.}\ \bibnamefont
  {Romalis}},\ }\bibfield  {title} {\bibinfo {title} {Comagnetometer probes of
  dark matter and new physics},\ }\href
  {https://doi.org/10.1088/2058-9565/ac1ae0} {\bibfield  {journal} {\bibinfo
  {journal} {Quantum Science and Technology}\ }\textbf {\bibinfo {volume}
  {7}},\ \bibinfo {pages} {014001} (\bibinfo {year} {2021})}\BibitemShut
  {NoStop}%
\bibitem [{\citenamefont {Lee}\ \emph {et~al.}(2018)\citenamefont {Lee},
  \citenamefont {Almasi},\ and\ \citenamefont {Romalis}}]{Lee2018}%
  \BibitemOpen
  \bibfield  {author} {\bibinfo {author} {\bibfnamefont {J.}~\bibnamefont
  {Lee}}, \bibinfo {author} {\bibfnamefont {A.}~\bibnamefont {Almasi}},\ and\
  \bibinfo {author} {\bibfnamefont {M.}~\bibnamefont {Romalis}},\ }\bibfield
  {title} {\bibinfo {title} {Improved limits on spin-mass interactions},\
  }\href {https://doi.org/10.1103/PhysRevLett.120.161801} {\bibfield  {journal}
  {\bibinfo  {journal} {Phys. Rev. Lett.}\ }\textbf {\bibinfo {volume} {120}},\
  \bibinfo {pages} {161801} (\bibinfo {year} {2018})}\BibitemShut {NoStop}%
\bibitem [{\citenamefont {Raffelt}(2012)}]{Raffelt2012}%
  \BibitemOpen
  \bibfield  {author} {\bibinfo {author} {\bibfnamefont {G.}~\bibnamefont
  {Raffelt}},\ }\bibfield  {title} {\bibinfo {title} {Limits on a
  {$CP$}-violating scalar axion-nucleon interaction},\ }\href
  {https://doi.org/10.1103/PhysRevD.86.015001} {\bibfield  {journal} {\bibinfo
  {journal} {Phys. Rev. D}\ }\textbf {\bibinfo {volume} {86}},\ \bibinfo
  {pages} {015001} (\bibinfo {year} {2012})}\BibitemShut {NoStop}%
\bibitem [{\citenamefont {Sheng}\ \emph {et~al.}(2014)\citenamefont {Sheng},
  \citenamefont {Kabcenell},\ and\ \citenamefont {Romalis}}]{sheng14a}%
  \BibitemOpen
  \bibfield  {author} {\bibinfo {author} {\bibfnamefont {D.}~\bibnamefont
  {Sheng}}, \bibinfo {author} {\bibfnamefont {A.}~\bibnamefont {Kabcenell}},\
  and\ \bibinfo {author} {\bibfnamefont {M.~V.}\ \bibnamefont {Romalis}},\
  }\bibfield  {title} {\bibinfo {title} {New classes of systematic effects in
  gas spin comagnetometers},\ }\href
  {https://doi.org/10.1103/PhysRevLett.113.163002} {\bibfield  {journal}
  {\bibinfo  {journal} {Phys. Rev. Lett.}\ }\textbf {\bibinfo {volume} {113}},\
  \bibinfo {pages} {163002} (\bibinfo {year} {2014})}\BibitemShut {NoStop}%
\bibitem [{\citenamefont {Feng}\ \emph {et~al.}(2020)\citenamefont {Feng},
  \citenamefont {Zhang}, \citenamefont {Lu},\ and\ \citenamefont
  {Sheng}}]{Feng2020}%
  \BibitemOpen
  \bibfield  {author} {\bibinfo {author} {\bibfnamefont {Y.-K.}\ \bibnamefont
  {Feng}}, \bibinfo {author} {\bibfnamefont {S.-B.}\ \bibnamefont {Zhang}},
  \bibinfo {author} {\bibfnamefont {Z.-T.}\ \bibnamefont {Lu}},\ and\ \bibinfo
  {author} {\bibfnamefont {D.}~\bibnamefont {Sheng}},\ }\bibfield  {title}
  {\bibinfo {title} {Electric quadrupole shifts of the precession frequencies
  of $^{131}\mathrm{Xe}$ atoms in rectangular cells},\ }\href
  {https://doi.org/10.1103/PhysRevA.102.043109} {\bibfield  {journal} {\bibinfo
   {journal} {Phys. Rev. A}\ }\textbf {\bibinfo {volume} {102}},\ \bibinfo
  {pages} {043109} (\bibinfo {year} {2020})}\BibitemShut {NoStop}%
\bibitem [{\citenamefont {Feng}\ \emph {et~al.}(2022)\citenamefont {Feng},
  \citenamefont {Ning}, \citenamefont {Zhang}, \citenamefont {Lu},\ and\
  \citenamefont {Sheng}}]{Feng2022}%
  \BibitemOpen
  \bibfield  {author} {\bibinfo {author} {\bibfnamefont {Y.-K.}\ \bibnamefont
  {Feng}}, \bibinfo {author} {\bibfnamefont {D.-H.}\ \bibnamefont {Ning}},
  \bibinfo {author} {\bibfnamefont {S.-B.}\ \bibnamefont {Zhang}}, \bibinfo
  {author} {\bibfnamefont {Z.-T.}\ \bibnamefont {Lu}},\ and\ \bibinfo {author}
  {\bibfnamefont {D.}~\bibnamefont {Sheng}},\ }\bibfield  {title} {\bibinfo
  {title} {Search for monopole-dipole interactions at the submillimeter range
  with a
  $^{129}\mathrm{Xe}\text{\ensuremath{-}}^{131}\mathrm{Xe}\text{\ensuremath{-}}\mathrm{Rb}$
  comagnetometer},\ }\href {https://doi.org/10.1103/PhysRevLett.128.231803}
  {\bibfield  {journal} {\bibinfo  {journal} {Phys. Rev. Lett.}\ }\textbf
  {\bibinfo {volume} {128}},\ \bibinfo {pages} {231803} (\bibinfo {year}
  {2022})}\BibitemShut {NoStop}%
\bibitem [{sup()}]{supp}%
  \BibitemOpen
  \href@noop {} {}\bibinfo {note} {See Sec. I of the Supplemental Material for
  a detailed description of the $^{129}$Xe-$^{131}$Xe-Rb comagnetometer, which
  includes Ref.~\cite{sheng14a,Feng2022}. See Sec. II for analyzing the nuclear
  spin precession signals, which includes Ref.~\cite{swallows2013}. See Sec.
  III for the effects of modulated Rb magnetometer, which includes
  Ref.~\cite{walker1997,Bulatowicz2013,Feng2022}. See Sec. III for the
  calculations of $\theta$ and related parameters. See Sec. IV for the
  calibration of $\beta$, which includes Ref.~\cite{cohen70}. See Sec. V for
  the studies of systematic effects, which includes
  Ref.~\cite{Feng2020}}\BibitemShut {NoStop}%
\bibitem [{\citenamefont {Walker}\ and\ \citenamefont
  {Larsen}(2016)}]{walker16}%
  \BibitemOpen
  \bibfield  {author} {\bibinfo {author} {\bibfnamefont {T.~G.}\ \bibnamefont
  {Walker}}\ and\ \bibinfo {author} {\bibfnamefont {M.~S.}\ \bibnamefont
  {Larsen}},\ }\bibfield  {title} {\bibinfo {title} {Spin-exchange-pumped {NMR}
  gyros},\ }\href@noop {} {\bibfield  {journal} {\bibinfo  {journal} {Advances
  in Atomic, Molecular, and Optical Physics}\ }\textbf {\bibinfo {volume}
  {65}},\ \bibinfo {pages} {373} (\bibinfo {year} {2016})}\BibitemShut
  {NoStop}%
\bibitem [{\citenamefont {Bulatowicz}\ \emph {et~al.}(2013)\citenamefont
  {Bulatowicz}, \citenamefont {Griffith}, \citenamefont {Larsen}, \citenamefont
  {Mirijanian}, \citenamefont {Fu}, \citenamefont {Smith}, \citenamefont
  {Snow}, \citenamefont {Yan},\ and\ \citenamefont {Walker}}]{Bulatowicz2013}%
  \BibitemOpen
  \bibfield  {author} {\bibinfo {author} {\bibfnamefont {M.}~\bibnamefont
  {Bulatowicz}}, \bibinfo {author} {\bibfnamefont {R.}~\bibnamefont
  {Griffith}}, \bibinfo {author} {\bibfnamefont {M.}~\bibnamefont {Larsen}},
  \bibinfo {author} {\bibfnamefont {J.}~\bibnamefont {Mirijanian}}, \bibinfo
  {author} {\bibfnamefont {C.~B.}\ \bibnamefont {Fu}}, \bibinfo {author}
  {\bibfnamefont {E.}~\bibnamefont {Smith}}, \bibinfo {author} {\bibfnamefont
  {W.~M.}\ \bibnamefont {Snow}}, \bibinfo {author} {\bibfnamefont
  {H.}~\bibnamefont {Yan}},\ and\ \bibinfo {author} {\bibfnamefont {T.~G.}\
  \bibnamefont {Walker}},\ }\bibfield  {title} {\bibinfo {title} {Laboratory
  search for a long-range {$T$}-odd, {$P$}-odd interaction from axionlike
  particles using dual-species nuclear magnetic resonance with polarized
  $^{129}\mathrm{Xe}$ and $^{131}\mathrm{Xe}$ gas},\ }\href
  {https://doi.org/10.1103/PhysRevLett.111.102001} {\bibfield  {journal}
  {\bibinfo  {journal} {Phys. Rev. Lett.}\ }\textbf {\bibinfo {volume} {111}},\
  \bibinfo {pages} {102001} (\bibinfo {year} {2013})}\BibitemShut {NoStop}%
\bibitem [{\citenamefont {Groten}(2000)}]{groten2000}%
  \BibitemOpen
  \bibfield  {author} {\bibinfo {author} {\bibfnamefont {E.}~\bibnamefont
  {Groten}},\ }\bibfield  {title} {\bibinfo {title} {Report of special
  commission 3 of {IAG}},\ }\href {https://doi.org/10.1017/S0252921100000488}
  {\bibfield  {journal} {\bibinfo  {journal} {International Astronomical Union
  Colloquium}\ }\textbf {\bibinfo {volume} {180}},\ \bibinfo {pages} {17}
  (\bibinfo {year} {2000})}\BibitemShut {NoStop}%
\bibitem [{\citenamefont {Schreiber}\ and\ \citenamefont
  {Wells}(2013)}]{schreiber2013}%
  \BibitemOpen
  \bibfield  {author} {\bibinfo {author} {\bibfnamefont {K.~U.}\ \bibnamefont
  {Schreiber}}\ and\ \bibinfo {author} {\bibfnamefont {J.-P.~R.}\ \bibnamefont
  {Wells}},\ }\bibfield  {title} {\bibinfo {title} {Invited review article:
  Large ring lasers for rotation sensing},\ }\href
  {https://doi.org/10.1063/1.4798216} {\bibfield  {journal} {\bibinfo
  {journal} {Rev. Sci. Instr.}\ }\textbf {\bibinfo {volume} {84}},\ \bibinfo
  {pages} {041101} (\bibinfo {year} {2013})}\BibitemShut {NoStop}%
\bibitem [{\citenamefont {Gemmel}\ \emph {et~al.}(2010)\citenamefont {Gemmel},
  \citenamefont {Heil}, \citenamefont {Karpuk}, \citenamefont {Lenz},
  \citenamefont {Ludwig}, \citenamefont {Sobolev}, \citenamefont {Tullney},
  \citenamefont {Burghoff}, \citenamefont {Kilian}, \citenamefont
  {Knappe-Gr$\ddot{u}$neberg}, \citenamefont {M$\ddot{u}$ller}, \citenamefont
  {Schnabel}, \citenamefont {Seifert}, \citenamefont {Trahms},\ and\
  \citenamefont {Bae${\beta}$ler}}]{gemmel2010}%
  \BibitemOpen
  \bibfield  {author} {\bibinfo {author} {\bibfnamefont {C.}~\bibnamefont
  {Gemmel}}, \bibinfo {author} {\bibfnamefont {W.}~\bibnamefont {Heil}},
  \bibinfo {author} {\bibfnamefont {S.}~\bibnamefont {Karpuk}}, \bibinfo
  {author} {\bibfnamefont {K.}~\bibnamefont {Lenz}}, \bibinfo {author}
  {\bibfnamefont {C.}~\bibnamefont {Ludwig}}, \bibinfo {author} {\bibfnamefont
  {Y.}~\bibnamefont {Sobolev}}, \bibinfo {author} {\bibfnamefont
  {K.}~\bibnamefont {Tullney}}, \bibinfo {author} {\bibfnamefont
  {M.}~\bibnamefont {Burghoff}}, \bibinfo {author} {\bibfnamefont
  {W.}~\bibnamefont {Kilian}}, \bibinfo {author} {\bibfnamefont
  {S.}~\bibnamefont {Knappe-Gr$\ddot{u}$neberg}}, \bibinfo {author}
  {\bibfnamefont {W.}~\bibnamefont {M$\ddot{u}$ller}}, \bibinfo {author}
  {\bibfnamefont {A.}~\bibnamefont {Schnabel}}, \bibinfo {author}
  {\bibfnamefont {F.}~\bibnamefont {Seifert}}, \bibinfo {author} {\bibfnamefont
  {L.}~\bibnamefont {Trahms}},\ and\ \bibinfo {author} {\bibfnamefont
  {S.}~\bibnamefont {Bae${\beta}$ler}},\ }\bibfield  {title} {\bibinfo {title}
  {Ultra-sensitive magnetometry based on free precession of nuclear spins},\
  }\href {https://doi.org/10.1140/epjd/e2010-00044-5} {\bibfield  {journal}
  {\bibinfo  {journal} {Eur. Phys. J. D}\ }\textbf {\bibinfo {volume} {57}},\
  \bibinfo {pages} {303} (\bibinfo {year} {2010})}\BibitemShut {NoStop}%
\bibitem [{\citenamefont {Dickerson}\ \emph {et~al.}(2013)\citenamefont
  {Dickerson}, \citenamefont {Hogan}, \citenamefont {Sugarbaker}, \citenamefont
  {Johnson},\ and\ \citenamefont {Kasevich}}]{dickerson2013}%
  \BibitemOpen
  \bibfield  {author} {\bibinfo {author} {\bibfnamefont {S.~M.}\ \bibnamefont
  {Dickerson}}, \bibinfo {author} {\bibfnamefont {J.~M.}\ \bibnamefont
  {Hogan}}, \bibinfo {author} {\bibfnamefont {A.}~\bibnamefont {Sugarbaker}},
  \bibinfo {author} {\bibfnamefont {D.~M.~S.}\ \bibnamefont {Johnson}},\ and\
  \bibinfo {author} {\bibfnamefont {M.~A.}\ \bibnamefont {Kasevich}},\
  }\bibfield  {title} {\bibinfo {title} {Multiaxis inertial sensing with
  long-time point source atom interferometry},\ }\href
  {https://doi.org/10.1103/PhysRevLett.111.083001} {\bibfield  {journal}
  {\bibinfo  {journal} {Phys. Rev. Lett.}\ }\textbf {\bibinfo {volume} {111}},\
  \bibinfo {pages} {083001} (\bibinfo {year} {2013})}\BibitemShut {NoStop}%
\bibitem [{\citenamefont {Yao}\ \emph {et~al.}(2018)\citenamefont {Yao},
  \citenamefont {Lu}, \citenamefont {Li}, \citenamefont {Luo}, \citenamefont
  {Wang},\ and\ \citenamefont {Zhan}}]{yao2018}%
  \BibitemOpen
  \bibfield  {author} {\bibinfo {author} {\bibfnamefont {Z.-W.}\ \bibnamefont
  {Yao}}, \bibinfo {author} {\bibfnamefont {S.-B.}\ \bibnamefont {Lu}},
  \bibinfo {author} {\bibfnamefont {R.-B.}\ \bibnamefont {Li}}, \bibinfo
  {author} {\bibfnamefont {J.}~\bibnamefont {Luo}}, \bibinfo {author}
  {\bibfnamefont {J.}~\bibnamefont {Wang}},\ and\ \bibinfo {author}
  {\bibfnamefont {M.-S.}\ \bibnamefont {Zhan}},\ }\bibfield  {title} {\bibinfo
  {title} {Calibration of atomic trajectories in a large-area
  dual-atom-interferometer gyroscope},\ }\href
  {https://doi.org/10.1103/PhysRevA.97.013620} {\bibfield  {journal} {\bibinfo
  {journal} {Phys. Rev. A}\ }\textbf {\bibinfo {volume} {97}},\ \bibinfo
  {pages} {013620} (\bibinfo {year} {2018})}\BibitemShut {NoStop}%
\bibitem [{\citenamefont {Schmidt}(1937)}]{Schmidt1937}%
  \BibitemOpen
  \bibfield  {author} {\bibinfo {author} {\bibfnamefont {T.}~\bibnamefont
  {Schmidt}},\ }\bibfield  {title} {\bibinfo {title} {Über die magnetischen
  momente der atomkerne},\ }\href {https://doi.org/10.1007/BF01338744}
  {\bibfield  {journal} {\bibinfo  {journal} {Zeitschrift für Physik}\
  }\textbf {\bibinfo {volume} {106}},\ \bibinfo {pages} {358} (\bibinfo {year}
  {1937})}\BibitemShut {NoStop}%
\bibitem [{\citenamefont {Kimball}(2015)}]{kimball2015}%
  \BibitemOpen
  \bibfield  {author} {\bibinfo {author} {\bibfnamefont {D.~F.~J.}\
  \bibnamefont {Kimball}},\ }\bibfield  {title} {\bibinfo {title} {Nuclear spin
  content and constraints on exotic spin-dependent couplings},\ }\href
  {https://doi.org/10.1088/1367-2630/17/7/073008} {\bibfield  {journal}
  {\bibinfo  {journal} {New Journal of Physics}\ }\textbf {\bibinfo {volume}
  {17}},\ \bibinfo {pages} {073008} (\bibinfo {year} {2015})}\BibitemShut
  {NoStop}%
\bibitem [{\citenamefont {Venema}(1994)}]{venemathesis}%
  \BibitemOpen
  \bibfield  {author} {\bibinfo {author} {\bibfnamefont {B.~J.}\ \bibnamefont
  {Venema}},\ }\href
  {http://www.pqdtcn.com/thesisDetails/7CFF2BB6F3176C4E42781CDFF03A3368}
  {\bibinfo {type} {Ph.d. thesis}},\ \bibinfo  {school} {University of
  Washington} (\bibinfo {year} {1994})\BibitemShut {NoStop}%
\bibitem [{\citenamefont {Tullney}\ \emph {et~al.}(2013)\citenamefont
  {Tullney}, \citenamefont {Allmendinger}, \citenamefont {Burghoff},
  \citenamefont {Heil}, \citenamefont {Karpuk}, \citenamefont {Kilian},
  \citenamefont {Knappe-Gr\"uneberg}, \citenamefont {M\"uller}, \citenamefont
  {Schmidt}, \citenamefont {Schnabel}, \citenamefont {Seifert}, \citenamefont
  {Sobolev},\ and\ \citenamefont {Trahms}}]{Tullney2013}%
  \BibitemOpen
  \bibfield  {author} {\bibinfo {author} {\bibfnamefont {K.}~\bibnamefont
  {Tullney}}, \bibinfo {author} {\bibfnamefont {F.}~\bibnamefont
  {Allmendinger}}, \bibinfo {author} {\bibfnamefont {M.}~\bibnamefont
  {Burghoff}}, \bibinfo {author} {\bibfnamefont {W.}~\bibnamefont {Heil}},
  \bibinfo {author} {\bibfnamefont {S.}~\bibnamefont {Karpuk}}, \bibinfo
  {author} {\bibfnamefont {W.}~\bibnamefont {Kilian}}, \bibinfo {author}
  {\bibfnamefont {S.}~\bibnamefont {Knappe-Gr\"uneberg}}, \bibinfo {author}
  {\bibfnamefont {W.}~\bibnamefont {M\"uller}}, \bibinfo {author}
  {\bibfnamefont {U.}~\bibnamefont {Schmidt}}, \bibinfo {author} {\bibfnamefont
  {A.}~\bibnamefont {Schnabel}}, \bibinfo {author} {\bibfnamefont
  {F.}~\bibnamefont {Seifert}}, \bibinfo {author} {\bibfnamefont
  {Y.}~\bibnamefont {Sobolev}},\ and\ \bibinfo {author} {\bibfnamefont
  {L.}~\bibnamefont {Trahms}},\ }\bibfield  {title} {\bibinfo {title}
  {Constraints on spin-dependent short-range interaction between nucleons},\
  }\href {https://doi.org/10.1103/PhysRevLett.111.100801} {\bibfield  {journal}
  {\bibinfo  {journal} {Phys. Rev. Lett.}\ }\textbf {\bibinfo {volume} {111}},\
  \bibinfo {pages} {100801} (\bibinfo {year} {2013})}\BibitemShut {NoStop}%
\bibitem [{mis()}]{miscRaffelt}%
  \BibitemOpen
  \href@noop {} {}\bibinfo {note} {Reference~\cite{Raffelt2012} only provides
  the astronomical limits of $g_s^Ng_p^N$, and we assume that $g_p^N=g_p^n$ in
  this case.}\BibitemShut {Stop}%
\bibitem [{\citenamefont {Krause}\ \emph {et~al.}(2023)\citenamefont {Krause},
  \citenamefont {Bertaux}, \citenamefont {McNamara}, \citenamefont {Gruenwald},
  \citenamefont {Longman}, \citenamefont {Scarlett},\ and\ \citenamefont
  {Fischbach}}]{krause2023}%
  \BibitemOpen
  \bibfield  {author} {\bibinfo {author} {\bibfnamefont {D.~E.}\ \bibnamefont
  {Krause}}, \bibinfo {author} {\bibfnamefont {J.}~\bibnamefont {Bertaux}},
  \bibinfo {author} {\bibfnamefont {A.~M.}\ \bibnamefont {McNamara}}, \bibinfo
  {author} {\bibfnamefont {J.~T.}\ \bibnamefont {Gruenwald}}, \bibinfo {author}
  {\bibfnamefont {A.}~\bibnamefont {Longman}}, \bibinfo {author} {\bibfnamefont
  {C.~Y.}\ \bibnamefont {Scarlett}},\ and\ \bibinfo {author} {\bibfnamefont
  {E.}~\bibnamefont {Fischbach}},\ }\bibfield  {title} {\bibinfo {title}
  {Phenomenological implications of a magnetic 5th force},\ }\href
  {https://doi.org/10.1142/S0217751X23500070} {\bibfield  {journal} {\bibinfo
  {journal} {International Journal of Modern Physics A}\ }\textbf {\bibinfo
  {volume} {38}},\ \bibinfo {pages} {2350007} (\bibinfo {year}
  {2023})}\BibitemShut {NoStop}%
\bibitem [{\citenamefont {Hunter}\ \emph {et~al.}(2013)\citenamefont {Hunter},
  \citenamefont {Gordon}, \citenamefont {Peck}, \citenamefont {Ang},\ and\
  \citenamefont {Lin}}]{Hunter2013}%
  \BibitemOpen
  \bibfield  {author} {\bibinfo {author} {\bibfnamefont {L.}~\bibnamefont
  {Hunter}}, \bibinfo {author} {\bibfnamefont {J.}~\bibnamefont {Gordon}},
  \bibinfo {author} {\bibfnamefont {S.}~\bibnamefont {Peck}}, \bibinfo {author}
  {\bibfnamefont {D.}~\bibnamefont {Ang}},\ and\ \bibinfo {author}
  {\bibfnamefont {J.-F.}\ \bibnamefont {Lin}},\ }\bibfield  {title} {\bibinfo
  {title} {Using the earth as a polarized electron source to search for
  long-range spin-spin interactions},\ }\href
  {https://doi.org/10.1126/science.1227460} {\bibfield  {journal} {\bibinfo
  {journal} {Science}\ }\textbf {\bibinfo {volume} {339}},\ \bibinfo {pages}
  {928} (\bibinfo {year} {2013})}\BibitemShut {NoStop}%
\bibitem [{\citenamefont {{Poddar}}\ and\ \citenamefont
  {{Pachhar}}(2023)}]{poddar2023}%
  \BibitemOpen
  \bibfield  {author} {\bibinfo {author} {\bibfnamefont {T.~K.}\ \bibnamefont
  {{Poddar}}}\ and\ \bibinfo {author} {\bibfnamefont {D.}~\bibnamefont
  {{Pachhar}}},\ }\bibfield  {title} {\bibinfo {title} {{Constraints on
  monopole-dipole potential from the tests of gravity: First bounds from single
  astrophysical observations}},\ }\bibfield  {journal} {\bibinfo  {journal}
  {arXiv e-prints}\ }\href {https://doi.org/10.48550/arXiv.2302.03882}
  {10.48550/arXiv.2302.03882} (\bibinfo {year} {2023})\BibitemShut {NoStop}%
\bibitem [{\citenamefont {Almasi}\ \emph {et~al.}(2020)\citenamefont {Almasi},
  \citenamefont {Lee}, \citenamefont {Winarto}, \citenamefont {Smiciklas},\
  and\ \citenamefont {Romalis}}]{Almasi2020}%
  \BibitemOpen
  \bibfield  {author} {\bibinfo {author} {\bibfnamefont {A.}~\bibnamefont
  {Almasi}}, \bibinfo {author} {\bibfnamefont {J.}~\bibnamefont {Lee}},
  \bibinfo {author} {\bibfnamefont {H.}~\bibnamefont {Winarto}}, \bibinfo
  {author} {\bibfnamefont {M.}~\bibnamefont {Smiciklas}},\ and\ \bibinfo
  {author} {\bibfnamefont {M.~V.}\ \bibnamefont {Romalis}},\ }\bibfield
  {title} {\bibinfo {title} {New limits on anomalous spin-spin interactions},\
  }\href {https://doi.org/10.1103/PhysRevLett.125.201802} {\bibfield  {journal}
  {\bibinfo  {journal} {Phys. Rev. Lett.}\ }\textbf {\bibinfo {volume} {125}},\
  \bibinfo {pages} {201802} (\bibinfo {year} {2020})}\BibitemShut {NoStop}%
\bibitem [{\citenamefont {Zhang}()}]{zhangprep}%
  \BibitemOpen
  \bibfield  {author} {\bibinfo {author} {\bibfnamefont {S.-B.}\ \bibnamefont
  {Zhang}},\ }\href@noop {} {}\bibinfo {howpublished} {in
  preparation}\BibitemShut {NoStop}%
\bibitem [{\citenamefont {Hao}\ \emph {et~al.}(2021)\citenamefont {Hao},
  \citenamefont {Yu}, \citenamefont {Yuan}, \citenamefont {Liu},\ and\
  \citenamefont {Sheng}}]{hao2021}%
  \BibitemOpen
  \bibfield  {author} {\bibinfo {author} {\bibfnamefont {C.-P.}\ \bibnamefont
  {Hao}}, \bibinfo {author} {\bibfnamefont {Q.-Q.}\ \bibnamefont {Yu}},
  \bibinfo {author} {\bibfnamefont {C.-Q.}\ \bibnamefont {Yuan}}, \bibinfo
  {author} {\bibfnamefont {S.-Q.}\ \bibnamefont {Liu}},\ and\ \bibinfo {author}
  {\bibfnamefont {D.}~\bibnamefont {Sheng}},\ }\bibfield  {title} {\bibinfo
  {title} {Herriott-cavity-assisted closed-loop {Xe} isotope comagnetometer},\
  }\href {https://doi.org/10.1103/PhysRevA.103.053523} {\bibfield  {journal}
  {\bibinfo  {journal} {Phys. Rev. A}\ }\textbf {\bibinfo {volume} {103}},\
  \bibinfo {pages} {053523} (\bibinfo {year} {2021})}\BibitemShut {NoStop}%
\bibitem [{\citenamefont {Swallows}\ \emph {et~al.}(2013)\citenamefont
  {Swallows}, \citenamefont {Loftus}, \citenamefont {Griffith}, \citenamefont
  {Heckel}, \citenamefont {Fortson},\ and\ \citenamefont
  {Romalis}}]{swallows2013}%
  \BibitemOpen
  \bibfield  {author} {\bibinfo {author} {\bibfnamefont {M.~D.}\ \bibnamefont
  {Swallows}}, \bibinfo {author} {\bibfnamefont {T.~H.}\ \bibnamefont
  {Loftus}}, \bibinfo {author} {\bibfnamefont {W.~C.}\ \bibnamefont
  {Griffith}}, \bibinfo {author} {\bibfnamefont {B.~R.}\ \bibnamefont
  {Heckel}}, \bibinfo {author} {\bibfnamefont {E.~N.}\ \bibnamefont
  {Fortson}},\ and\ \bibinfo {author} {\bibfnamefont {M.~V.}\ \bibnamefont
  {Romalis}},\ }\bibfield  {title} {\bibinfo {title} {Techniques used to search
  for a permanent electric dipole moment of the $^{199}${Hg} atom and the
  implications for $\mathit{CP}$ violation},\ }\href
  {https://doi.org/10.1103/PhysRevA.87.012102} {\bibfield  {journal} {\bibinfo
  {journal} {Phys. Rev. A}\ }\textbf {\bibinfo {volume} {87}},\ \bibinfo
  {pages} {012102} (\bibinfo {year} {2013})}\BibitemShut {NoStop}%
\bibitem [{\citenamefont {Walker}\ and\ \citenamefont
  {Happer}(1997)}]{walker1997}%
  \BibitemOpen
  \bibfield  {author} {\bibinfo {author} {\bibfnamefont {T.~G.}\ \bibnamefont
  {Walker}}\ and\ \bibinfo {author} {\bibfnamefont {W.}~\bibnamefont
  {Happer}},\ }\bibfield  {title} {\bibinfo {title} {Spin-exchange optical
  pumping of noble-gas nuclei},\ }\href {https://doi.org/DOI
  10.1103/RevModPhys.69.629} {\bibfield  {journal} {\bibinfo  {journal} {Rev.
  Mod. Phys.}\ }\textbf {\bibinfo {volume} {69}},\ \bibinfo {pages} {629}
  (\bibinfo {year} {1997})}\BibitemShut {NoStop}%
\bibitem [{\citenamefont {Cohen-Tannoudji}\ \emph {et~al.}(1970)\citenamefont
  {Cohen-Tannoudji}, \citenamefont {Dupont-Roc}, \citenamefont {Haroche},\ and\
  \citenamefont {Lalo{\"e}}}]{cohen70}%
  \BibitemOpen
  \bibfield  {author} {\bibinfo {author} {\bibfnamefont {C.}~\bibnamefont
  {Cohen-Tannoudji}}, \bibinfo {author} {\bibfnamefont {J.}~\bibnamefont
  {Dupont-Roc}}, \bibinfo {author} {\bibfnamefont {S.}~\bibnamefont
  {Haroche}},\ and\ \bibinfo {author} {\bibfnamefont {F.}~\bibnamefont
  {Lalo{\"e}}},\ }\bibfield  {title} {\bibinfo {title} {Diverses r{\'e}sonances
  de croisement de niveaux sur des atomes pomp{\'e}s optiquement en champ nul.
  i. th{\'e}orie},\ }\href@noop {} {\bibfield  {journal} {\bibinfo  {journal}
  {Rev. Phys. Appl.}\ }\textbf {\bibinfo {volume} {5}},\ \bibinfo {pages} {95}
  (\bibinfo {year} {1970})}\BibitemShut {NoStop}%
\end{thebibliography}
\end{document}